# Response to Comment on "Low-frequency lattice phonons in halide perovskites explain high defect tolerance toward electron-hole recombination"

Weibin Chu[1, 2], Qijing Zheng[1], Oleg V. Prezhdo[2*], Jin Zhao[1,3,4*] and Wissam A. Saidi[5*]

[1] ICQD/Hefei National Laboratory for Physical Sciences at the Microscale, CAS Key Laboratory of Strongly-Coupled Quantum Matter Physics, and Department of Physics, University of Science and Technology of China, Hefei, Anhui 230026, China

[2] Departments of Chemistry, and Physics and Astronomy, University of Southern California, Los Angeles, CA 90089, United States

[3] Department of Physics and Astronomy, University of Pittsburgh, Pittsburgh PA 15260, United States

[4] Synergetic Innovation Center of Quantum Information & Quantum Physics, University of Science and Technology of China, Hefei, Anhui 230026, China

[5] Department of Mechanical Engineering and Materials Science, University of Pittsburgh, Pittsburgh, Pennsylvania 15261, United States

* Email: alsaidi@pitt.edu; zhaojin@ustc.edu.cn; prezhdo@usc.edu

**Abstract**

Recently we proposed that defect tolerance in the hybrid perovskites is due to their characteristic low-frequency lattice phonon modes that decrease the non-adiabatic coupling and weaken the overlap between the free carrier and defect states [Sci. Adv. 6 7, eaaw7453 (2020)]. Kim and Walsh disagree with the interpretation and argue that there are flaws in our employed methodology. Herein we address their concerns and show that their conclusions are not valid due to misunderstandings of nonadiabatic transition.

In our recent study, we proposed that defect tolerance in the hybrid perovskites originates from the characteristic low-frequency lattice phonon modes of the system, which lead to small non-adiabatic coupling and weaken the overlap between the free carrier and defect states (*1*). Kim and Walsh raised concerns over several methodological factors, and questioned the veracity of our conclusions (*2*). Here, we show that their arguments are not valid and are due to misinterpreting factors related to electron-phonon coupling and non-adiabatic molecular dynamic simulations. Further, we argue that they overlooked several important and carefully analyzed aspects of our study as detailed below. In conclusion, we do not see any substantial influence of their concerns on our reported findings.

**Nonadiabatic transitions vs. state crossings.** Kim and Walsh argued that the nonradiative electron-hole (*e-h*) recombination is not physical if there is no band gap collapse between conduction band minimum (CBM) and valence band maximum (VBM). We disagree with this statement. Nonadiabatic transitions in different systems, in which the crossing of the electronic states is not necessary, have been investigated by NAMD simulation for many years (*3-9*). For example, as illustrated in ref. (*10*), nonadiabatic transitions can happen in three prototype regimes, i.e., via the Landau-Zener, nonadiabatic tunneling and Rosen-Zener processes. In the Landau-Zener and nonadiabatic tunneling processes, there are energy states crossings in the diabatic space. The adiabatic levels do not cross each other due to the avoided-crossing principle. Particularly, in the nonadiabatic tunneling processes, a significant barrier can exist between the two potential energy surfaces. Therefore, the band gap collapse suggested by Kim and Walsh does not apply. Moreover, even if there are no crossings in the diabatic space, as described in the Rosen-Zener type of transition, the nonadiabatic transition can still take place (*9-12*). As discussed in our study (*1*), the nonradiative *e-h* recombination rate has a positive correlation with the nonadiabatic coupling elements (NAC), and NAC is inversely proportional to the band gap as shown in the equation 1.

$$d_{jk} = \langle \psi_j | \nabla_R | \psi_k \rangle \cdot \dot{R} = \frac{\langle \psi_j | \nabla_R \hat{H} | \psi_k \rangle}{\varepsilon_k - \varepsilon_j} \dot{R} \qquad (1)$$

Therefore, in a wide band gap semiconductor, NAC is very small, but non zero, suggesting that electron-hole (*e-h*) recombination is not completely forbidden and can rarely take place. Such a nonradiative process affects solar energy conversion and photon luminescence in different systems, including both molecules and solids, which have been intensively studied (*5, 6, 13, 14*). In many of these nonradiative processes, there are no electronic state crossings or the systems remains far from such crossings.

**Shockley-Read-Hall model in different semiconductors.** Based on our previous studies on MAPbI$_3$, CsPbI$_3$ and TiO$_2$ (*1, 15, 16*), we propose that the Shockley-Read-Hall (SRH) model works for many conventional semiconductors because the deep band gap states can introduce an additional *e-h* recombination pathway, and the excess charge can significantly increase the electron-phonon (*e-ph*) coupling. Our previous study of TiO$_2$ clearly showed that the deep defect level can form an *e-h* recombination center. For this system, the NAC is increased by one order of magnitude when TiO$_2$ is doped with a Cr-N pair, due to the reduction of band gap and the increase of *e-ph* coupling, $\langle \psi_j | \nabla_R \hat{H} | \psi_k \rangle$. This suggests that SRH works well in TiO$_2$. Similar results are found in MoS$_2$ (*14*). In addition, our preliminary unpublished work on GaAs also suggests the validity of SRH. However, MAPbI$_3$ is a special material for several reasons as argued in our study (*1*). The NAC between the CBM and VBM of the pristine system is as small as 0.69 meV, that is only half of the corresponding TiO$_2$ value. More importantly, native defects in MAPbI$_3$ are found to further reduce the NACs. We explained this counterintuitive behavior in MAPbI$_3$ due to the characteristic low-frequency lattice phonon modes. First, low-frequency phonon suggests small nuclear velocity that enters the definition of the NAC, see Eq. (1). Thus, even with defects, the NACs are always small. Second, soft lattice will induce charge localization, which will again reduce the NAC, contrary to the TiO$_2$ system. Thus, defect-assisted-recombination is hindered due to the reduced NAC compared to conventional semiconductors. This peculiar behavior cannot be predicted by the SRH model due to the lack of an explicit electron-phonon coupling in this model.

**Calculation of the NAC using all-electron wavefunctions**. Kim and Walsh argued that the NACs are overestimated since we are using projector-augmented-wave (PAW) pseudo-wavefunctions instead of all-electron (AE) wavefunctions. To show the implication of this, **Kim and Walsh** computed the overlap $\langle\psi_j|\psi_k\rangle$ between two different Kohn-Sham (KS) orbitals for a single atomic structure using both methods and showed that $\langle\psi_j|\psi_k\rangle = \delta_{jk}$ for AE wavefunctions but not for the pseudo-wavefunction for the case of GaAs. However, as we show below, differences between the pseudo or AE wavefunctions is strongly misinterpreted.

First, we noticed that the pseudo-wavefunctions that **Kim and Walsh** used are not normalized. Thus, the wavefunction overlap cannot be the identity matrix. However, normalization is part of our standard protocol for the calculation of the NAC using pseudo-wavefunctions. As we show in Table 1, the normalized pseudo-wavefunctions already give a good description of the overlap. Additionally, one can fully re-orthonormalize the pseudo-wavefunctions. However, it makes little difference in the NAC calculation, as we detail in Table 2.

Second, the overalp $\langle\psi_j|\psi_k\rangle$ that **Kim and Walsh** used is not related to the NAC, which is computed as wavefunction overlap between *different* atomic configurations and not at the *same* atomic configuration. As we discuss below, pseudo-wavefunctions provide a very good approximation to the NAC due to a cancellation effect.

The NAC matrix element $d_{jk}$ is computed in our simulations numerically as ref. (*17*):

$$d_{jk} = \left\langle\psi_j\left|\frac{\partial}{\partial t}\right|\psi_k\right\rangle = \frac{1}{2dt}\begin{pmatrix} \langle\psi_j(r,t)|\psi_k(r,t+dt)\rangle - \langle\psi_j(r,t)|\psi_k(r,t)\rangle \\ +\langle\psi_j(r,t+dt)|\psi_k(r,t+dt)\rangle - \langle\psi_j(r,t+dt)|\psi_k(r,t)\rangle \end{pmatrix} \quad (2)$$

**Table 1**. Calculated overlap between the Kohn-Sham wavefunctions of CsPbI$_3$ with iodine vacancy at the Γ point. Calculations are performed with the pseudo-wavefunction from VASP (a) without normalization as reported in Ref. (*2*), (b) after normalization as used in our calculations, and (c) after orthonormalization.

| (a) | Pseudo-Wavefunction (non-normalized) | | | (b) | Pseudo-Wavefunction (normalized) | | | (c) | Pseudo-Wavefunction (orthonormalized) | | |
|---|---|---|---|---|---|---|---|---|---|---|---|
| | VBM | Defect | CBM | | VBM | Defect | CBM | | VBM | Defect | CBM |
| **VBM** | 1.22 | 0.02 | 0.01 | **VBM** | 1 | 0.02 | 0.01 | **VBM** | 1 | 0 | 0 |
| **Defect** | 0.02 | 1.08 | 0.01 | **Defect** | 0.02 | 1 | 0.00 | **Defect** | 0 | 1 | 0 |
| **CBM** | 0.01 | 0.01 | 1.10 | **CBM** | 0.01 | 0.00 | 1 | **CBM** | 0 | 0 | 1 |

That is, the NAC can be understood as the difference in the overlap of states $\psi_j$ and $\psi_k$ of different atomic configurations at consecutive $t$ and $t + \Delta t$ time steps. The NAC measures response of the wavefunctions to atomic motion, and this is why it is associated with *e-ph* coupling. A small nuclear velocity associated with low-frequency phonons induces small NAC, because atom displacements between $t$ and $t + \Delta t$ are small. Note that the 2$^{nd}$ and 3$^{rd}$ terms in the above expression are zero for orthogonal wavefunctions, and they are not present in the original work (17). These terms that correct for the nonorthogonality and give accurate NAC as shown in Table 2 are present in our calculations.

The proper calculation of the overlap $\langle \psi_j | \psi_k \rangle$ using AE wavefunctions contains four terms,

$$\langle \psi_j | \psi_k \rangle = O_0 + O_1 + O_2 + O_3 \tag{3}$$

where

$$O_0 = \langle \tilde{\psi}_j | \tilde{\psi}_k \rangle \tag{4}$$

is the overlap of the pseudo-wavefunctions $\tilde{\psi}_j$ and $\tilde{\psi}_k$, and

$$O_1 = \sum_m \langle \tilde{\psi}_j | \tilde{p}_m \rangle (\langle \phi_m | - \langle \tilde{\phi}_m |) | \tilde{\psi}_k \rangle) \tag{5}$$

$$O_2 = \sum_n \langle \tilde{\psi}_j | (|\phi_n\rangle - |\tilde{\phi}_n\rangle) \langle \tilde{p}_n | \tilde{\psi}_k \rangle \tag{6}$$

$$O_3 = \sum_m \sum_n \langle \tilde{\psi}_j | \tilde{p}_m \rangle (\langle \phi_m | - \langle \tilde{\phi}_m |)(|\phi_n\rangle - |\tilde{\phi}_n\rangle) \langle \tilde{p}_n | \tilde{\psi}_k \rangle \tag{7}$$

are the contributions from the core part and the projection (*18*). We used PAWPYSEED (*18*) to evaluate the core part, as Kim and Walsh reported in their comment. We carried out NAMD simulations for the iodine vacancy ($I_V$) defect in $CsPbI_3$ using NAC computed using both AE and pseudo-wavefunctions. The NAC calculated with the pseudo-wavefunction was reported in ref. (*15*). The comparison with the AE wavefunction is shown in Figure 1. The NACs of a representative configuration are also provided in Table 2. The data show that the pseudo-wavefunction and the AE wavefunctions yield almost the same results, suggesting the core part contribution can be neglected. In general, the core contributions ($O_1$, $O_2$ and $O_3$) do not appreciably change between $t$ and $t + \Delta t$. In addition, the 2$^{nd}$ and 3$^{rd}$ terms of Eq. (2) correct them for non-orthogonality. Thus, the contributions of the core regions of the AE wavefunction to the NAC are negligible.

**Table 2**. Calculated NAC between the Kohn-Sham wavefunctions of $CsPbI_3$ with iodine vacancy at the Γ point. Calculations are performed with the pseudo-wavefunction from VASP with (a) normalization, as reported in our published manuscript, and (b) orthonormalization. (c) Calculations were performed with the AE wavefunction.

| (a) | Pseudo-Wavefunction (normalized) | | | (b) | Pseudo-Wavefunction (orthonormalized) | | | (c) | AE Wavefunction | | |
|---|---|---|---|---|---|---|---|---|---|---|---|
| | VBM | Defect | CBM | | VBM | Defect | CBM | | VBM | Defect | CBM |
| **VBM** | 0 | 0.10 | 0.32 | **VBM** | 0 | 0.11 | 0.32 | **VBM** | 0 | 0.09 | 0.34 |
| **Defect** | 0.10 | 0 | 0.96 | **Defect** | 0.10 | 0 | 0.95 | **Defect** | 0.09 | 0 | 0.93 |
| **CBM** | 0.32 | 0.96 | 0 | **CBM** | 0.32 | 0.95 | 0 | **CBM** | 0.34 | 0.93 | 0 |

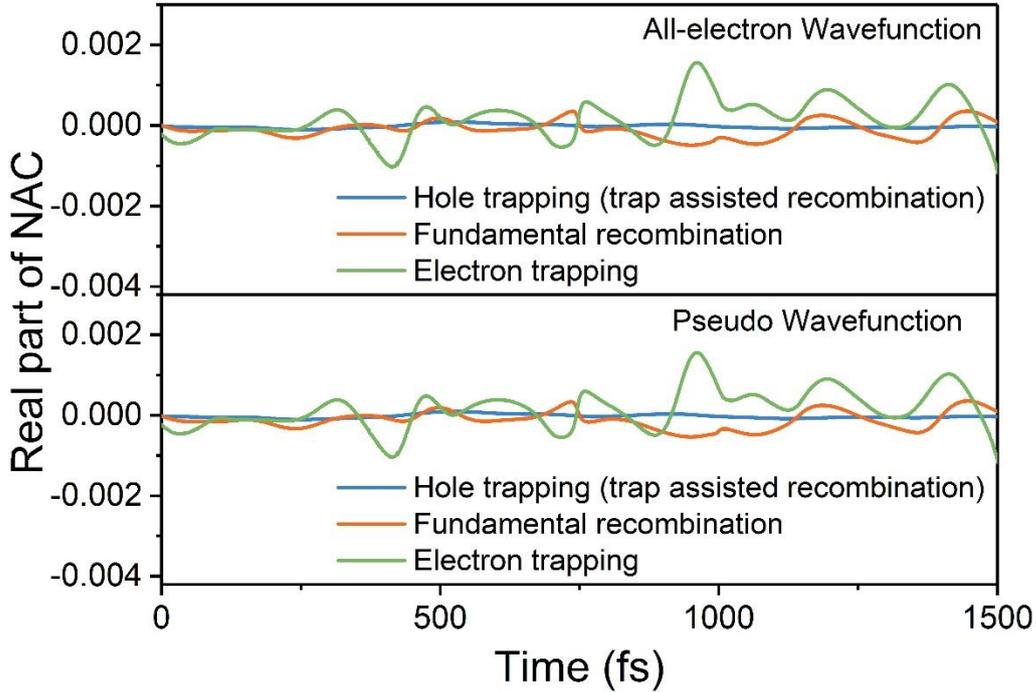

**Fig. 1** The real part of the NAC calculated using the pseudo-wavefunction and the all-electron wavefunction.

Third, for completeness, we noted that the phase correction has to be accounted for to correctly compute the NAC. The adiabatic wavefunctions from different atomic configurations can differ by an arbitrary phase, or sign for real-valued wavefunction, because they are solutions to the eigenvalue problem that is invariant with respect to wavefunction phase. Thus, the overlap of the KS orbitals from different configurations is not uniquely determined. This is often referred to as the "phase consistency" problem (*19*). Our NAC code keeps track of the KS orbitals phase.

**Gamma point sampling**. Kim and Walsh stated that "… In addition, the band-edge states of semiconductors will be poorly described due to Γ-point sampling of the Brillouin zone." We agree with this statement. However, we are puzzled about the relevance of this statement for our study. All of our calculations are done using a 4×4×2 k-point grid as clearly stated in the MATERIALS AND METHODS part.

**The effect of supercell size simulation timescale**. Kim and Walsh argued that the supercell size employed in our study is not large enough and the simulation timescale is not long enough. We agree with these concerns, and we have already discussed them in our paper (1). The finite supercell size can affect the *e-h* recombination time, since the electron and the hole are limited in the supercell, which will increase the likelihood of their recombination. As we discussed in (1), when the supercell is increased from primitive to 2×2×2, the e-h recombination timescale increases from 11 to 64 ns for the pristine MAPbI3. Therefore we expect longer lifetime with larger supercells that may be closer to the experimental value. In addition, other long range effects such as ferroelectricity and large polaron may further suppress the e-h recombination. Importantly, we showed in the manuscript that both $MA_I$ and $I_i$ suppress the e-h recombination, which is in good agreement with the results of the 1×1×1 unit cell. **This suggests that the conclusions drawn in our study are not sensitive to the size of the unit cell.**

**The charged system**. Kim and Walsh argued that we had not investigated the charged defect systems. We agree that in Ref. (*1*) we only investigated neutral defect systems. The charged systems should be an interesting topic for future investigations. However, our main argument on defect tolerance due to low-frequency phonon modes for the neutral systems is likely to be valid for the charged systems as well, if, similarly to the neutral systems, the CBM, VBM and defect states are only coupled with low-frequency phonons. In this case, these low-frequency phonons are responsible for the nonradiative charge trapping and recombination in the charged systems.

**Fractional decay in dynamics**. Kim and Walsh argued that "Even if the proper matrix elements were employed, the kinetics of trap-mediated recombination are first-order and limited by the capture of the minority carrier. For a positive trap, the excess electronic energy is emitted as heat at each sequential discrete capture events. In contrast, NAMD simulates a bimolecular electron-hole excitation of a defective unit cell and monitors the *continuous* fractional decay of an excited state after a finite simulation time…". We see no contrast here. The simulations mimic the experimental process as closely as possible. In the experiment, absorption of a photon creates an electron-hole pair. Such electron-hole pair constitutes the initial state of our NAMD simulation. Either charge can be trapped independently of the presence of the other charge, and the trapping probability is first order in the charge concentration. The excess electronic energy is emitted as heat in NAMD.

According to the fundamental interpretation of quantum mechanics, the continuous fractional decay of the excited state population represented by the square of the corresponding wavefunction coefficient is interpreted as probability of a discrete event. Moreover, the surface hopping framework (*20*) is constructed exactly to represent such discrete events. NAMD trajectories contain discrete trapping and recombination events, and the data are averaged over many trajectories, representing an ensemble of systems, as measured experimentally.

Most rate expressions, such as Fermi's golden rule, are based on perturbation theories. In comparison, the transition probabilities in NAMD simulations are calculated from the solution of the time-dependent Schrödinger equation and are not perturbative. **Mono-molecular recombination, bi-molecular recombination as well as high-order Auger type recombination can be simulated well with NAMD**, as demonstrated previously. Each trajectory is independent of the other ones. For instance, in one trajectory, a carrier can be trapped by the defect and can stay trapped until recombined with a free carrier, or it can be de-trapped without further recombination. In another trajectory, the electron and hole can recombine directly without been trapped. All of these processes are considered simultaneously in our time-dependent NAMD simulations. This is explained in the 3$^{rd}$ paragraph of the results part in Ref. (*1*). Thus, discrete capture events and heat emission in trap-mediated recombination are intrinsically considered in NAMD in a real-time manner.

**Conclusion**. We would like to express our sincere gratitude to Kim and Walsh for bringing up the above points. The NAMD calculations are still state-of-the-art, and many new users are not aware of the technical details that are routinely used by NAMD developers and seasoned users. We believe the systematic investigations on the *e-h* recombination rate in different semiconductors with various defects and impurities are very important. Until now, we have investigated TiO$_2$, black phosphorus and the perovskite solar cell (*1, 15, 16, 21*). Other groups have investigated transition metal dichalcogenides (*5, 14*). There are many other semiconductors that need to be investigated and understood. We strongly believe that *ab initio* NAMD simulation provides a powerful tool to capture all these unique phenomena beyond what is offered by the simple SRH model.